# Clouds in Partial Atmospheres of Lava Planets and Where to Find Them

T. Giang Nguyen[1], Nicolas B. Cowan[1,2], and Lisa Dang[3]  
[1] Department of Physics, McGill University, 3600 rue University, Montréal, QC H3A 2T8, Canada  
[2] Department of Earth & Planetary Sciences, 3450 rue University, Montréal, QC H3A 0E8, Canada  
[3] Trottier Institute for Research on Exoplanets and Department of Physics, Université de Montréal, 1375 Ave. Thérèse-Lavoie-Roux, Montréal, QC H2V 0B3, Canada

## Abstract

With dayside temperatures hot enough to sustain a magma ocean and a silicate atmosphere, lava planets are the best targets for studying the atmosphere of a rocky world. In the absence of nightside heating, the entire atmosphere collapses near the day–night terminator, so condensation seems inevitable, but the impact of clouds on radiative transfer, dynamics, and observables has not yet been studied in the nonglobal atmospheric regime. Therefore, we simulate cloud formation and determine which lava planets should be most affected by clouds. We find that despite the scattering of visible light by clouds, heat advection compensates for the cooling effect of clouds in the atmosphere. On the other hand, surface temperatures are significantly affected and can drop by 100–200 K under a cloudy sky. We find that among our targets, HD213885b and HD20329b are most affected by cloud formation: there is a discernible difference between having clouds and not having them, but the precision required to make such an inference is at the limit of current instruments.

*Unified Astronomy Thesaurus concepts:* Exoplanet atmospheres (487)

## 1. Introduction

### 1.1. Background

Hundreds of Earth-sized planets have been discovered (E. R. Adams et al. 2021), but inferring their surface and atmospheric properties remains a significant challenge. Our best options for the in-depth characterization of rocky planets are currently ultrashort-period planets. Phase curves obtained with the Spitzer Space Telescope have enabled tentative inferences of atmospheres on the highly irradiated super-Earths K2-141b (S. Zieba et al. 2022) and 55 Cnc-e (B.-O. Demory et al. 2016a, 2016b; S. J. Mercier et al. 2022).

A more powerful instrument is needed for definitive atmospheric detection and characterization of thin silicate atmospheres. Indeed, the recently launched James Webb Space Telescope (JWST) has observed the ultrashort-period planets K2-141b (L. Dang et al. 2021; N. Espinoza et al. 2021) and 55 Cnc-e (R. Hu et al. 2024) and will observe TOI-561b (J. Teske et al. 2023) and K2-22b (J. T. Wright et al. 2023). Such observations can provide unique clues about terrestrial planet evolution.

Ultrashort-period rocky planets with dayside temperatures >1800 K are theorized to harbor permanent dayside magma oceans and are often referred to as lava planets. Since the silicate melt is directly exposed to the atmosphere, atmospheric dynamics are intrinsically dependent on interior dynamics (E. S. Kite et al. 2016; C. Dorn & T. Lichtenberg 2021; C.-É. Boukaré et al. 2022). Magma oceans are a common result of planetary formation (L. Schaefer & L. T. Elkins-Tanton 2018) and play a critical role in sequestering volatiles (K. Moore et al. 2023), so lava planets provide a unique window into the early solar system and Earth (M. M. Hirschmann 2012; K. J. Zahnle et al. 2020). For a complete overview of silicate outgassing and the corresponding chemistry and radiative-convective dynamics, see R. Wordsworth & L. Kreidberg (2022).

Tidally locked planets risk having their atmosphere condense out entirely on the permanent nightside (R. Wordsworth 2015). However, tidally locked terrestrial planets may sustain a global atmosphere if the initial outgassed atmosphere is thick enough (M. Joshi 2003). Therefore, general circulation models (GCMs) are adequate for simulating a thick-atmosphere regime for lava planets such as 55 Cnc-e (M. Hammond & R. T. Pierrehumbert 2017). A general source of atmospheric circulation on tidally locked planets can be found in R. T. Pierrehumbert & M. Hammond (2019). Despite such robust work on lava planet atmospheres, the nonglobal atmospheric regime is less studied and cannot be simulated with a GCM. Two approaches are therefore used to simulate a thin lava planet atmosphere: either focusing on the dayside radiative transfer or the hydrodynamics.

Radiative transfer simulations of lava planets show a silicate atmosphere with a strong vertical temperature inversion (e.g., Y. Ito et al. 2015; M. Zilinskas et al. 2022; A. A. Piette et al. 2023). These studies often find the atmosphere hotter than the surface, as heat cannot be advected to cooler regions. On the other hand, hydrodynamical models of nonglobal atmospheres often simplify, if not outright ignore, radiative transfer (e.g., T. Castan & K. Menou 2011; T. G. Nguyen et al. 2020). However, this method can trace the horizontal heat and mass transport due to day–night winds. Both approaches ultimately lack self-consistent cloud formation, which affects both the fluid dynamics and radiative transfer.

### 1.2. Objective

In anticipation of JWST observations of lava planets, we simulate cloud formation and quantify its effect on thin-atmospheric dynamics and observations. Although there are many published works specific to exoplanet clouds (for a review, see C. Helling 2019), it is impossible to use a sophisticated cloud model within an existing nonglobal atmosphere modeling framework. We therefore use a simple parameterization of cloud formation consistent with the turbulent boundary layer model of

T. G. Nguyen et al. (2022), which is the only published work that has coupled radiative transfer with hydrodynamical processes for nonglobal atmospheres.

We use this model to simulate the atmospheres of five lava planets: K2-141b (L. Malavolta et al. 2018), TOI-431b (A. Osborn et al. 2021), HD20329b (F. Murgas et al. 2022), HD213885b (N. Espinoza et al. 2020), and 55 Cnc-e (J. N. Winn et al. 2011; but we note that its recently published eclipse spectrum suggests a volatile-full planet—R. Hu et al. 2024). The selected candidates have some of the best signal-to-noise ratios for rocky exoplanets and they will likely be observed extensively. Therefore, we compare these lava planets to predict where cloud formation might be expected.

## 2. Methods

### 2.1. The Hydrodynamical Model

Thin and tenuous atmospheres are common on moons and dwarf planets of the outer solar system. Huge temperature fluctuations on these bodies induce periods and locations of atmospheric collapse and massive surface outgassing, e.g., on Io (C. C. Tsang et al. 2016) or Pluto (G. R. Gladstone & L. A. Young 2019). Due to the stark changes in the fluid regime, it is difficult to implement GCMs and hydrodynamics are best approximated using boundary layer analysis (A. P. Ingersoll et al. 1985). Because the atmospheres of lava planets are expected to behave similarly, the same calculations are useful, if not necessary.

The expectation of synchronous rotation due to tidal locking makes lava planets easier to deal with than solar system objects. Having a permanent dayside and nightside allows for steady-state flow. Due to the complexities in handling nonlinear terms around evaporation and condensation, we model a pure atmosphere. We are also forced to neglect atmospheric escape, which should happen for lighter volatiles (Y. Ito & M. Ikoma 2021), but is justified if we track heavy molecules like SiO. Finally, we neglect the Coriolis force so as to preserve the axial symmetry of tidally locked objects. Generally, for planets with rotational periods of multiple days, the kilometer per second wind speeds are strong enough to make the Coriolis force insignificant (T. Castan & K. Menou 2011). However, our targets have a period of ⩽1 day and so we acknowledge that the Coriolis force may play a large role in dynamics, but implementing it is beyond our current capability.

By applying the approximations described above, we can construct a framework that describes the flow along the planetary boundary layer. The formulation of the base evaporation-driven flow is akin to the shallow-water equations of A. P. Ingersoll et al. (1985). The equations essentially calculate pressure, winds, and the temperature of the atmosphere at half the scale height. A detailed formulation can be found in Appendix A.1.

### 2.2. Radiative Transfer

Our implementation of radiative transfer largely follows T. G. Nguyen et al. (2022), and a detailed formulation can be found in Appendix A.2. T. G. Nguyen et al. (2022) split up the radiative transfer into three bands (UV, visible, and IR), with an averaged absorption cross section within each band. To improve on this scheme, we now calculate radiative absorption and emission line by line in the spectrum, doing away with averaging spectral bands. We validate our radiative transfer scheme by directly comparing our wind-free silicate atmosphere with M. Zilinskas et al. (2022), as shown in Appendix A.3. We find that our atmospheric temperatures match well with theirs (within 10%), whereas the old scheme of T. G. Nguyen et al. (2022) can be off by up to 25%, especially for hotter and brighter stars.

We adopt the absorption cross section of SiO from S. N. Yurchenko et al. (2022). As for the stellar spectra, we found proxy analogs in the MUSCLES treasury surveys for K2-141, TOI-431, and 55 Cnc-e: their closest analogs are HD85512, HD40307, and HD97658, respectively (K. France et al. 2016). For our targets of HD213885 and HD20329, their properties are closest to the Sun and so we use the solar spectrum; the spectra are shown in the bottom panel of Figure 1. The absolute magnitude of each analog's spectrum is scaled to the target's effective stellar temperature and orbital distance.

### 2.3. Cloud Implementation

Clouds are abundant on exoplanets, but their modeling and detection have largely been confined to hot Jupiters (B.-O. Demory et al. 2013) and sub-Neptunes (L. Kreidberg et al. 2014). Some of these clouds likely consist of silicate material like SiO and $MgSiO_3$ (D. Powell et al. 2018; P. Gao et al. 2021). Although lava planets have exotic atmospheric dynamics, their cloud composition is not, at least by the standards of exoplanets. Cloud particles on lava planets may be analogous to dust found on catastrophically evaporating rocky worlds (A. Curry et al. 2024).

Rapid mass exchange between the surface and the atmosphere of a lava planet may lead to meteorological processes driven by the recycling of rocky material between the surface and atmosphere (C.-É. Boukaré et al. 2022). Volatile components like Na and K may form clouds due to interactions with stellar winds, but these light materials are subjected to atmospheric loss (Y. Ito & M. Ikoma 2021) or freeze-out in nightside alkali glaciers (T. G. Nguyen et al. 2020).

Because we are focusing on lava planets with Earth-like bulk composition (excluding 55 Cnc-e), we can reasonably assume that volatiles such as water or $CO_2$ are absent from the atmosphere (see R. J. Graham et al. 2021). Silicate condensates remain the dominant constituent, but if the dew point is too close to the surface, the silicate vapor may not condense midair but simply deposit directly on the surface of the magma ocean (L. Schaefer & B. Fegley 2009). We do not know how much of the SiO vapor will condense into cloud particles, nor do we know the specific properties of these particles. Since clouds might have significant impact on the atmosphere of a lava planet, ignoring them altogether is not a viable option. We therefore opt to parameterize cloud effects into our hydrodynamical model in the same spirit as how radiative transfer was implemented, i.e., as simply as possible.

As cloud formation is inherently tied to the condensation rate $C$, we define a cloud optical depth proportional to $C$: $\tau_c = f_c\, C$. The dimensionless $f_c$ parameterizes the fraction of condensible material that actually turns into cloud particles as well as the particle properties that affect optical depth, such as particle size and density. Although much nuance is lost in this formulation, current observations only allow us to probe the temperature of the atmosphere and surface. Therefore, our main aim is to estimate to what extent clouds affect the overall thermodynamics and see if clouds can produce a detectable signal.

By creating and testing a range of $f_c$, we are essentially inferring how the planet behaves under varying strength of cloud coverage. This helps to identify important aspects, such as minimum detectable cloud formation, which planets are most affected by clouds, and direct inference of cloud properties.

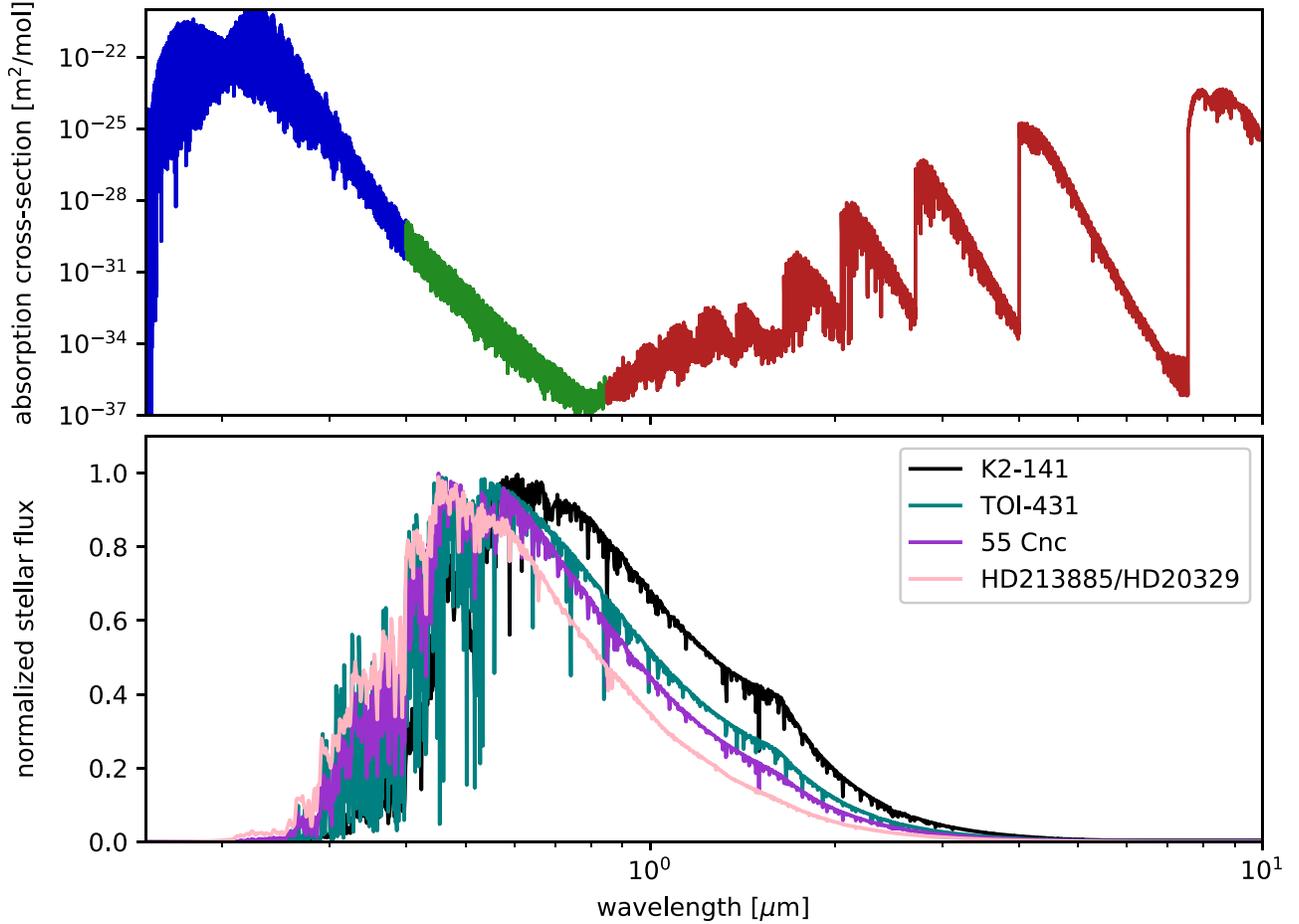

**Figure 1.** Top: the absorption cross section of SiO (S. N. Yurchenko et al. 2022). The blue, green, and red show the UV, visible, and IR ranges. Bottom: the normalized flux of each star using close analogs.

We calculate the resultant opacity ($\epsilon_c$) following Equation (A17). Here, we make another simplifying cloud assumption: cloud particles scatter visible light isotropically, so that half the incident visible stellar light ($F_{*,\mathrm{vis}}$) is reflected upward when the cloud opacity is 1. This assumes that the clouds are transparent in the IR and that most of the UV is absorbed by the upper levels of the atmosphere. Note that ignoring cloud absorption in the IR neglects the greenhouse effect, removing significant complexities to the system of equations. This yields $F_c$, the stellar flux that is reflected because of clouds:

$$F_c = \frac{1}{2}\,\epsilon_c\,F_{*,\mathrm{vis}}. \quad (1)$$

Because SiO is nearly transparent at the visible wavelengths (Figure 1), the atmospheric energy balance is unaffected by $F_c$. However, the surface energy balance is significantly affected, as visible light is absorbed entirely by the surface. This stems from treating the surface as a blackbody but is justified in the expected low geometric albedo of magma surfaces (Z. Essack et al. 2020): the surface energy is calculated from the bulk bolometric flux minus stellar radiation absorbed by the atmosphere plus the emissions from the atmosphere minus the visible stellar flux that is reflected by clouds minus latent heat when evaporating. The surface energy balance then becomes

$$\sigma T_s = F_* - F_{\mathrm{stel}} + F_{\mathrm{RC}} - F_c - Q_{\mathrm{lat}}, \quad (2)$$

where $\sigma$ is the Stefan–Boltzmann constant. The surface temperature, $T_s$, depends on the overlying atmospheric properties and vice versa (i.e., Equations (A4) and (A9)), coupling the surface and atmosphere. $F_*$ is the bulk stellar flux, $F_{\mathrm{stel}}$ is the stellar flux that has been absorbed by the atmosphere, $F_{\mathrm{RC}}$ is the thermal emission from the atmosphere, $F_c$ is the stellar flux that was reflected by clouds, and $Q_{\mathrm{lat}}$ is the latent heat of vaporization.

### 2.4. Simulated Observations

This subsection will describe how the state variables $P$, $V$, and $T$, computed as in Appendix A.1, are used to simulate observations. Because a lava planet's atmosphere is so thin and does not extend much past the terminator, transmission spectroscopy of the gravitationally bound atmosphere is challenging (T. G. Nguyen et al. 2022). We therefore focus on planetary emission. We first calculate the top-of-atmosphere outgoing radiative flux $F_{\mathrm{TOA}}$:

$$F_{\mathrm{TOA}}(\lambda) = \epsilon_\lambda B(\lambda, T) + (1 - \epsilon_\lambda)B(\lambda, T_s), \quad (3)$$

where the first term is atmospheric emission and the second is surface emission. We take into account the adiabatic profile

**Table 1**
System Parameters Used for the Hydrodynamical Model

| Planet Name | Radius ($R_E$) | Stellar Radius ($R_*$) | Distance ($a/R_*$) | Stellar Temp. (K) | $g$ (ms$^{-2}$) | Source |
| --- | --- | --- | --- | --- | --- | --- |
| K2-141b | 1.54 | 0.681 | 2.30 | 4373 | 21.8 | O. Barragán et al. (2018) |
| TOI-431b | 1.28 | 0.731 | 3.32 | 4850 | 18.4 | A. Osborn et al. (2021) |
| HD20329b | 1.72 | 1.13 | 3.42 | 5596 | 24.6 | F. Murgas et al. (2022) |
| HD213885b | 1.74 | 1.10 | 3.93 | 5978 | 28.5 | N. Espinoza et al. (2020) |
| 55 Cnc-e | 1.87 | 0.964 | 3.52 | 5317 | 22.3 | V. Bourrier et al. (2018) |

**Note.** Although these values do not show up in the equations mentioned in the previous section, with the exception of $g$, the orbital properties are needed to calculate the incident stellar flux pattern, $F_*(\theta)$, following Z. Kopal (1954).

where $T(P)$ is calculated at the pressure where the optical depth is 2/3. For any location $\theta$, the flux can be calculated and so we convert temperature maps into flux maps.

We convert $F_{\mathrm{TOA}}(\theta)$ into a map of colatitude ($\alpha$) and longitude ($\phi$), assuming polar symmetry about the substellar point. We then integrate the flux across the visible hemisphere, following N. B. Cowan & E. Agol (2008):

$$F_e(\lambda, \zeta) = \int_0^\pi \int_{\zeta-\pi/2}^{\zeta+\pi/2} F(\alpha, \phi, \lambda) \sin^2(\alpha) \cos(\phi - \zeta) d\phi d\alpha, \quad (4)$$

where $F_e$ is the emission signal and $\zeta$ is the position of the planet in its orbit ($\zeta = 0$ at superior conjunction and $\zeta = \pi$ at inferior conjunction). Keeping $\lambda$ constant and varying $\zeta$ between 0 and $2\pi$ yields a phase curve at wavelength $\lambda$. Keeping $\zeta$ at 0 while varying $\lambda$ yields the emission spectrum at eclipse.

## 3. Results

### 3.1. Output from Model

All the equations from the previous section together form a self-consistent model that couples the atmosphere and surface along with rudimentary cloud formation. To simulate our specific targets, we apply the planetary parameters listed in Table 1.

We show the initial results without any cloud formation in Figure 2. We see that atmospheric pressure is directly related to surface temperature: the hotter the surface, the thicker the atmosphere. However, despite having cooler surface temperatures, HD213885b and HD20329b have much hotter atmospheres; they also have stronger winds.

We plot the contributions of all radiative terms in Figure 3. With the exception of K2-141b, most of the atmospheric absorption is in the UV, owing to the absorption features of SiO in that range. Most of the radiative cooling occurs in the IR, although HD213885b and HD20329b are hot enough to emit significant UV radiation.

Despite the enormous magnitude of the radiative fluxes, incoming and outgoing radiation tend to reach equilibrium: the net radiative balance is 1–2 orders of magnitude smaller than each individual radiative term.

### 3.2. Results with Clouds

To review, we have parameterized cloud optical effects using the parameter $f_c$ in Section 2. Note that our simple implementation of clouds neglects certain feedback, such as the greenhouse effect of clouds. With the assumption that cloud formation must be proportional to the condensation rate $C$, the optical depth and opacity must subsequently also depend on $C$. Therefore, $f_c$ acts as a substitute to estimate how much of the condensible material becomes cloud, as well as cloud properties such as droplet size; we explore these properties in Section 4.

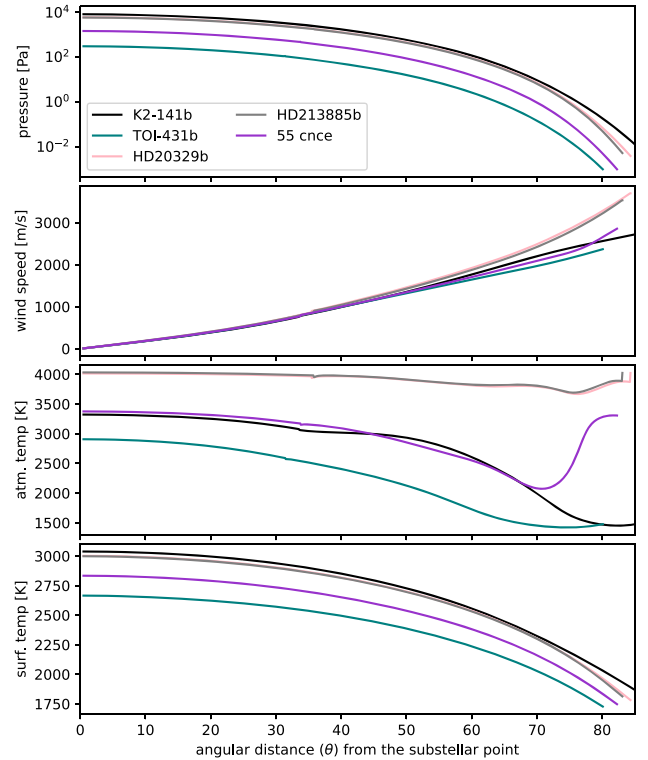

**Figure 2.** State variables half a scale height above the surface for selected lava planets. From top to bottom, the panels show pressure, wind speed, atmospheric temperature, and surface temperature. The solutions end when the flow no longer behaves as a continuous fluid and the calculations break down; this is essentially where the horizontal mass advection reaches near 0 and the atmospheric collapse is complete. K2-141b has the thickest atmosphere, as it receives the most stellar flux and therefore has the hottest surface temperature. However, K2-141b's atmosphere is significantly cooler than HD213885b and HD20329b, as atmospheric temperature is largely dependent on stellar UV flux. The temperature jump for 55 Cnc-e at the edge of the atmosphere can be attributed to a dynamical effect described by A. P. Ingersoll et al. (1985), where the air viscosity begins to induce significant heating at certain pressure and temperature levels; the same phenomenon can be seen in T. Castan & K. Menou (2011).

We restrict $f_c$ between $10^2$ and $10^4$; any lower and cloud effects are negligible across all planets, while anything higher will make the opacity uniformly 1. Because the SiO vapor absorption is weak in the visible, clouds do not directly affect the radiative budget of the atmosphere but do so via atmosphere–surface interactions. The biggest impact, therefore, is on the surface energy budget, as expected from our formulation (see Equation (2)). We show the condensation rate, optical depth, opacity, and surface temperature of cloudy simulations in Figure 4.

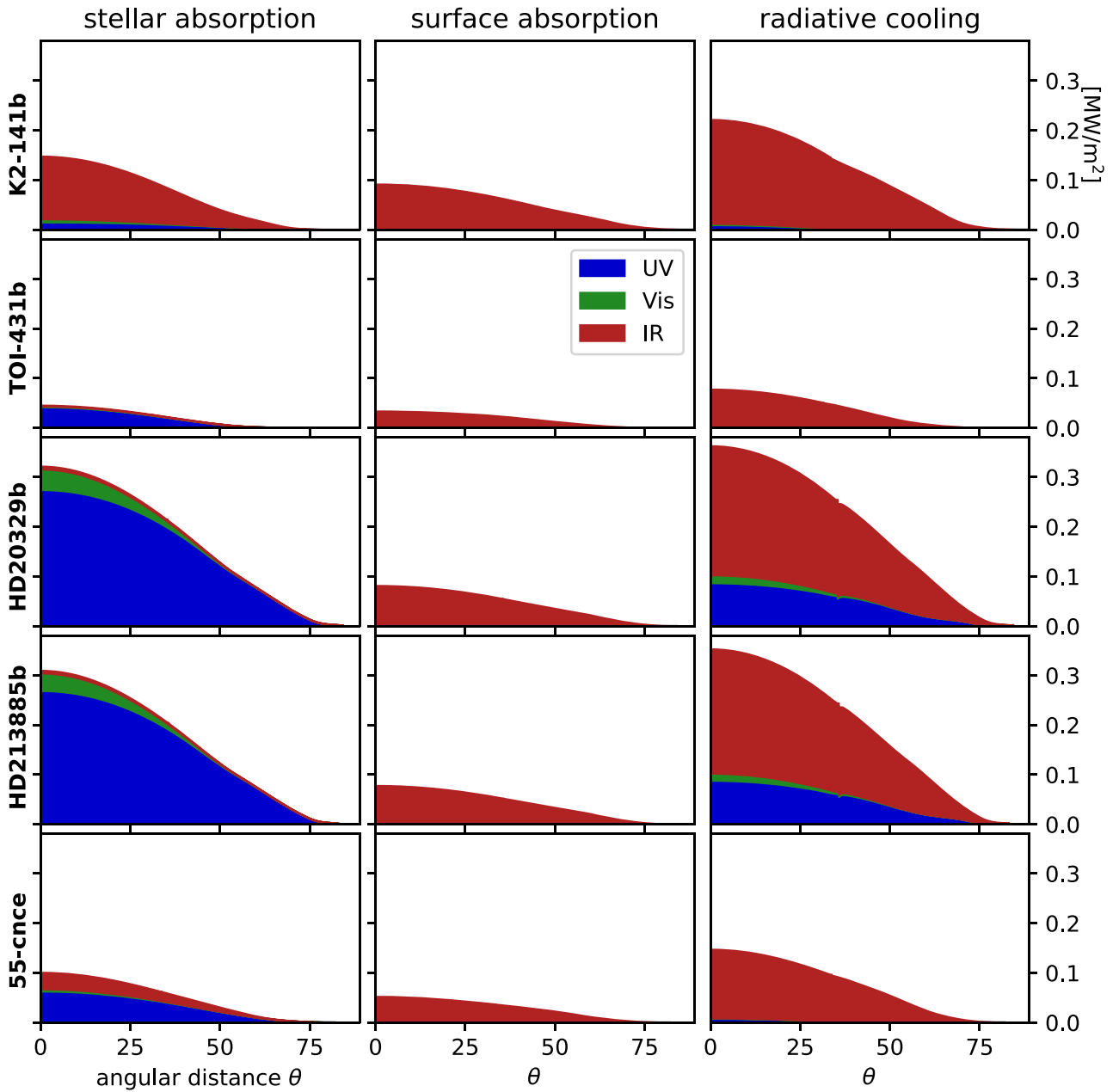

**Figure 3.** Radiative budget of atmosphere of our five lava planets. The first column shows the absorption of stellar flux, the second shows the absorption of blackbody radiation from the surface, and the third shows radiative cooling. K2-141 is less UV-bright than the other stars, so the stellar absorption of its planet is mostly in the IR. Because of the high UV stellar absorption on HD213885b and HD20329b, their atmospheres become hot enough to emit significant UV radiation.

### 3.3. Simulated Observations

Given a global map for surface and atmospheric temperature, we calculate the outgoing radiative flux at the top of the atmosphere using Equation (3). We fix the time at eclipse and integrate the flux over the visible hemisphere using Equation (4) at different wavelengths $\lambda$ to yield Figure 5.

## 4. Discussion

### 4.1. On Radiative Energy Balance

We see that the stellar spectrum predictably dictates how hot the atmosphere gets. Despite the surface of K2-141b receiving more bolometric stellar flux than HD20329b and HD213885b, its atmosphere absorbs less stellar flux, as the star K2-141 emits mostly in the visible and IR. Meanwhile, HD213885b and HD20329b absorb more stellar UV and visible radiation, thus providing more heat to their atmospheres.

The coarser radiative scheme of T. G. Nguyen et al. (2022) tended to overestimate the UV radiative balance while underestimating the IR. With a finer spectral resolution, we see that UV emission never dominates and that the SiO spectral

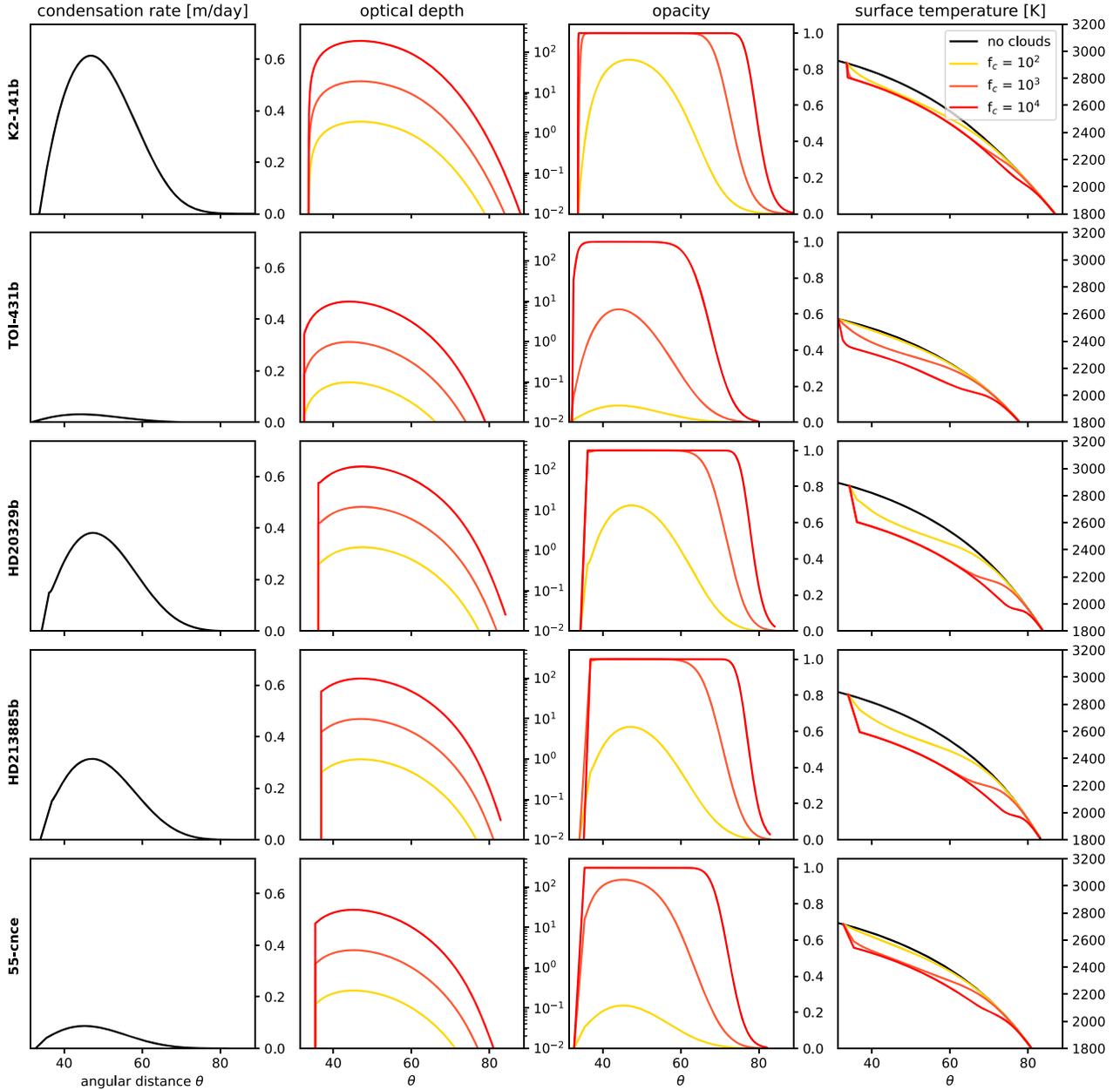

**Figure 4.** The impact of clouds on lava planet surface temperature. The first column is the condensation rate. We zoom in on $\theta$ where condensation occurs. We impose the factors $f_c$ onto the condensation rate to create the optical depth plots shown in the second column, and the opacity is plotted in the third column. The final column is the surface temperature below the cloud deck. Clouds cool the surface significantly (up to 200 K), but the changes in atmospheric temperature are an order of magnitude smaller.

features in the IR are strong enough to emit flux even when the pressure is low.

### 4.2. Vertical Temperature Profile

1D radiative-convective models predict that while the lower silicate atmosphere of lava planets does follow an adiabatic profile, there is a sharp temperature inversion at the top of the atmosphere, due to shallow UV penetration in the atmosphere (M. Zilinskas et al. 2022; A. A. Piette et al. 2023). Hydrodynamical studies, on the other hand, predict that advection of heat weakens this inversion (T. G. Nguyen et al. 2022).

Although we have adopted an adiabatic vertical temperature profile, we can guess what happens when the vertical temperature lapse rate is inverted. This can be done by increasing the $\beta$ value (this parameter is calculated from vertical pressure integration in Equation (A12)). T. G. Nguyen et al. (2022) showed that arbitrarily increasing $\beta$ leads to stronger horizontal flow but an overall cooler atmosphere. Within the calculations, raising $\beta$ increases the horizontal pressure gradient force, thus increasing the wind velocity, $V$. However, the energy balance does not change, so any gains in the kinetic term, $V^2/2$, translate to a loss in the internal heat term, $CpT$ (see Equation (A3)). The physical reasoning is that because the atmospheric inversion inhibits mixing, horizontal winds are significantly increased, but since energy must be conserved, the air's internal heat energy must be lower.

When balancing the incoming radiation in the atmosphere, we find that stellar absorption is comparable to surface blackbody absorption for K2-141b, TOI-431b, and 55 Cnc-e; this suggests a rather isothermal $T$–$P$ profile, as there is as much heat from the top as there is at the bottom. However, for HD213885b and HD20329b, there is significantly more stellar absorption than surface absorption. With significantly more heating from the top, most of which is UV flux, we would expect a strong temperature inversion, as predicted by M. Zilinskas et al. (2022). While we have kept a consistent vertical temperature profile in this study to isolate the effects of clouds, we note that future simulations of lava planets should aspire to use self-consistent vertical profiles, given the spectrum of the host star.

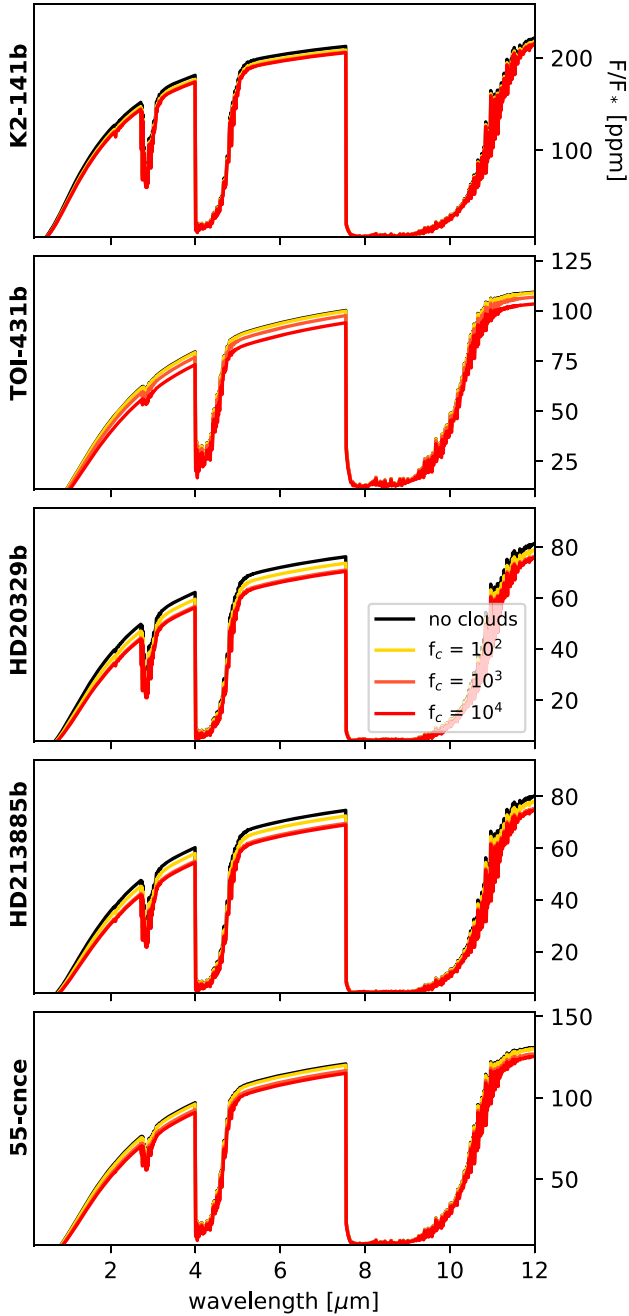

**Figure 5.** Simulated eclipse spectra of selected planets. The spectrum is calculated from the top-of-the-atmosphere emission, which ultimately depends on the temperatures of both the atmosphere and the surface. In strong SiO spectral features, emission originates from the upper layers of the atmosphere. For an adiabatic $T$–$P$ profile, this leads to an extremely cold temperature reading and thus the eclipse spectra show deep absorption features. Although clouds do cool the atmosphere slightly, the changes in the eclipse depth are barely visible. Since clouds cool the surface more, the overall blackbody curve of the spectra is lowered and the difference between cloudy and cloudless is seen relatively most clearly on HD20329b and HD213885.

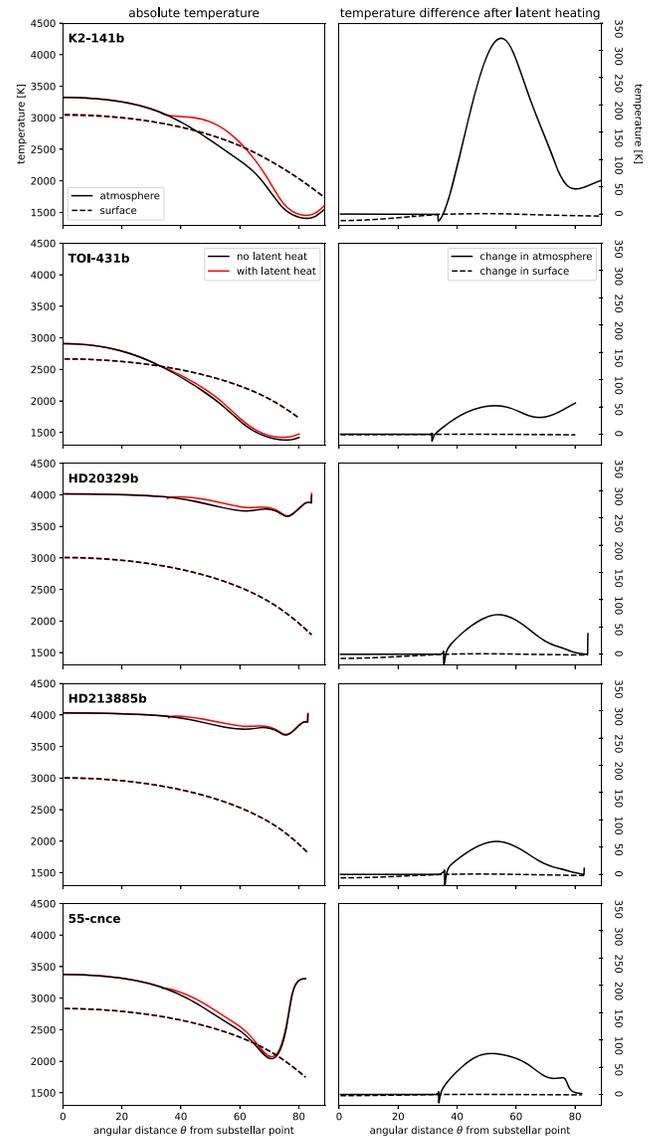

**Figure 6.** We show the changes in atmospheric and surface temperature when latent heat is implemented self-consistently in our code. Latent heat cools the surface when evaporating and warms the atmosphere when condensing, because we assume that condensation occurs aloft. As predicted by E. S. Kite et al. (2016), temperature changes due to latent heating are negligible at the surface but warm the atmosphere a little. However, the changes are minor, with the exception of K2-141b, where atmospheric temperatures can warm by 10%. This is due to the relatively large condensation rate but relatively low radiative heating and cooling caused by orbiting around a cooler star.

### 4.3. Latent and Chemical Heat

As the radiative terms tend to cancel out, the net energy budget of the atmosphere is controlled by dynamics terms, such as latent heat and sensible heat. We show in Figure 6 that the latent heat is negligible, as predicted by E. S. Kite et al. (2016). Even when coupling the latent heat with the atmosphere where radiative transfer is hampered by the absorption cross section, latent heating changes the atmosphere by less than 5%. However, K2-141b has a relatively larger condensation rate and a relatively lower radiative absorptivity due to its cooler star; temperatures can increase up to 10% when latent heating is accounted for. Sensible heating is still the main driver of the flow and it is composed of heat flux from two directions: horizontally, via flow-induced heat advection, and vertically, via temperature mixing between the surface and atmosphere.

Although the latent heat is relatively small, the vaporization of the surface and subsequent condensation is not a simple phase change. The magma ocean is principally composed of $SiO_2$ (assuming a bulk silicate Earth composition; L. Schaefer & B. Fegley 2009), while the atmosphere contains mostly SiO, hence evaporation and condensation involve the dissociation and recombination of oxygen atoms. In the context of hot gaseous planets, T. J. Bell & N. B. Cowan (2018) showed that the energy of the dissociation and recombination of hydrogen molecules can be significant. We estimate the energy of the reaction $SiO_2 \Rightarrow SiO + (1/2)O_2$ based on the difference in the enthalpy of formation between SiO and $SiO_2$ (the enthalpy of $O_2$ is 0). With this difference being 811 kJ mol$^{-1}$ (M. W. Chase & US 1998), the energy per unit mass of SiO is $2 \times 10^7$ J kg$^{-1}$, an order of magnitude greater than the latent heat of vaporization ($10^6$ J kg$^{-1}$; E. S. Kite et al. 2016). Unlike the latent heat, this chemical energy may dominate the energy budget, but we leave this problem to future studies, since our current model can only accommodate a single gas species.

### 4.4. The Effects of Clouds on Lava Planet Climate

The formation of clouds has a weaker effect on the atmosphere than on the surface, due to the fact that the atmosphere is optically thin at visible wavelengths, even in the absence of clouds. Visible light reflected by clouds would have mostly passed through the atmosphere in any case; it is the surface that loses that energy and thus can cool by 200 K.

Smaller $f_c$ correspond to larger cloud particles, which are radiatively less important. Assuming spherical droplets, we can decompose $f_c$ into cloud particle properties via

$$f_c = \frac{3}{8} \frac{t}{\rho \, d} p_e. \quad (5)$$

The density, $\rho$, of the condensed $SiO_2$ is 2650 kg m$^{-3}$ and $p_e$ is the fraction of condensible material that forms cloud droplets instead of condensing directly on the surface of the magma ocean—this depends on the altitude at which clouds form. The particle diameter, $d$, typically ranges from submicrons to tens of microns for silicate clouds in hot-Jupiter atmospheres (D. Powell et al. 2018), while dust grain analysis of catastrophically evaporating rocky worlds yields 0.1–1 μm (J. Budaj 2013). Finally, $t$ is the timescale over which the cloud droplet stays in the atmosphere. Two timescales could be at play: the time it takes for the cloud droplet to fall to the surface or the time it takes for the cloud droplet to be advected from the point of nucleation to where the atmosphere collapses.

We estimate the terminal velocity of free-falling cloud particles falling through 2500 K gas under an acceleration $g = 25$ m s$^{-2}$ using Stokes' law: terminal velocity, $V = (2/9)(\rho_c - \rho_a)gr^2/\eta$, where $\rho_c$ is the particle density and $\rho_a$ is the air density. With micron-size particles, the terminal velocity is a very slow $2 \times 10^{-4}$ m s$^{-1}$. We surmise that the horizontal advection timescale is likely the limiting factor. It takes $\sim$5000 s to transport a parcel of gas 45° of angular distance for a planetary radius of 1.5 $R_E$ and a wind speed of 1.5 km s$^{-1}$. Therefore, we get $f_c \sim 10^4$, given $d = 10^{-6}$ m and $p_e = 0.1$.

The effect of clouds on the atmospheric energy budget is minimal because we have assumed clouds are transparent in the IR, but the surface is influenced significantly. Because we have treated the magma ocean as a still surface, it does not have any mechanism to advect heat (see C.-É. Boukaré et al. 2022). Therefore, the surface is entirely dependent on instellation, and losing up to half of all visible light due to clouds results in significant temperature decreases that might be detectable via astronomical observations.

In the simulated eclipse spectra shown in Figure 4, we see that the greater the cloud opacity, the lower the overall emission. Clouds have the biggest impacts on the dynamics of HD213885b and HD20329b, due to the high incident stellar radiation at visible wavelengths. However, this difference in overall signal is too small for current instruments to precisely characterize. Ultimately, K2-141b remains the best target lava planet for atmospheric detection because of the signal-to-noise ratio, but it is the worst target for differentiating between cloudy and clear-sky conditions.

## 5. Conclusions

Clouds are a fundamental part of a planet's meteorology. Lava planets are most likely to have, at minimum, nonglobal atmospheres. In this regime, clouds should appear as the atmosphere starts to condense out. Therefore, we have simulated the atmospheres of the lava planets K2-141b, TOI-431b, HD20329b, HD213885b, and 55 Cnc-e to see which of these planets are best for cloud detection and characterization.

With or without clouds, our calculations show that atmospheric temperature is more dependent on the stellar spectrum than on bolometric flux. HD213885b and HD20329b, whose stars emit more UV light, have hotter atmospheres than K2-141b, despite receiving less instellation.

We implement clouds by scaling the optical depth to the condensation rate using the $f_c$ parameter to roughly approximate weak, medium, and strong cloud effects. We find that atmospheric properties are not significantly affected by visible light scattering of clouds: the existing heat advection from the evaporating region through sensible heat as well as latent heat is enough to offset clouds. The surface is more affected by clouds and the effect is predictably stronger for greater optical instellation. As such, the overall effect of clouds is best seen on HD213885b and HD20329b but is still challenging to distinguish.

Implementing an adiabatic profile up to the top of the atmosphere creates deep absorption features that render the spectroscopic signature of clouds indistinguishable; adopting the silicate T–P profiles of M. Zilinskas et al. (2022) and A. A. Piette et al. (2023) should yield more accurate observables, which may reveal an opportunity to detect clouds, but the results of such 1D models must be reconciled with atmospheric dynamics.

### Acknowledgments

This research was partially funded by the Canadian Space Agency's James Webb Space Telescope observer's program and the Heising-Simons Foundation. N.B.C. acknowledges support from an NSERC Discovery Grant, a Tier 2 Canada Research Chair, and an Arthur B. McDonald Fellowship. The authors also thank the Trottier Space Institute and l'Institut de recherche sur les exoplanètes for their financial support and dynamic intellectual environment.

# Appendix

## A.1. Base Hydrodynamical Formulation

The framework for the hydrodynamical calculation is a system akin to the shallow-water equations describing the conservation of mass (Equation (A1)), momentum (Equation (A2)), and energy (Equation (A3)), following A. P. Ingersoll et al. (1985):

$$\partial_x(\rho h V) = m E, \tag{A1}$$

$$\partial_x(\rho h V^2) = -\partial_x \int_z P dz + \tau, \tag{A2}$$

$$\partial_x\left(\rho h V\left(\frac{V^2}{2} + C_p T\right)\right) = Q_{\text{sens}} + Q_{\text{lat}} + Q_{\text{RT}}, \tag{A3}$$

where $x$ and $z$ denote the horizontal and vertical direction coordinates. $P$, $V$, and $T$ are the state variables pressure, velocity, and temperature. The variables $\rho$ and $h$ are air density and column thickness. The parameters $m$ and $C_p$ are the molecular mass and heat capacity, respectively. Equation (A1) describes how the change of material moving along the flow is controlled by the mass exchange between the surface and the atmosphere via evaporation, $E$. Equation (A2) describes how the momentum of the air parcel is controlled by the horizontal pressure gradient force and surface drag, $\tau$. Equation (A3) describes how the kinetic energy and internal heat energy of the parcel are controlled by the energy fluxes, which include sensible ($Q_{\text{sens}}$), latent ($Q_{\text{lat}}$), and radiative heating terms ($Q_{\text{RC}}$) and are expanded upon later in the text.

The mass flux, $E$ [molecules s$^{-1}$ m$^{-2}$], is the term that describes what is added and what is removed from the atmosphere. This is controlled by evaporation:

$$E = \frac{P - P_v(T_s)}{m\sqrt{2\pi R T_s}}, \tag{A4}$$

where $P_v$ is the saturated vapor pressure taken from Y. Miguel et al. (2011) and $T_s$ is the surface temperature. $R$ is the gas constant. To further calculate the fluxes, such as the surface drag $\tau$ and components of $Q$, we define a mean flow velocity $V_e = mE/\rho$ and an eddy velocity $V_d = V_*^2/V$. The frictional velocity $V_*$ is calculated iteratively from the implicit relation with $V$:

$$V = 2.5 V_* \log\left(\frac{9.0 \, V_* \, H\rho}{2\eta}\right), \tag{A5}$$

where $H$ is the scale height, $\rho$ is the density, and $\eta$ is the temperature-dependent viscosity.[4] With these velocities, we further define advection coefficients that describe how heat and momentum are dynamically transferred between the surface and atmosphere, $w_a$ and $w_s$:

$$w_a = \frac{2V_d^2}{V_e + 2V_d}, \qquad (V_e > 0) \tag{A6a}$$

$$w_s = \frac{V_e^2 + 2V_d V_e + 2V_d^2}{V_e + 2V_d}, \tag{A6b}$$

---
[4] $\eta = 1.8 \times 10^{-5}(T/291)^{3/2}(411)/(T+120)$, following Sutherland's ideal gas formula (T. Castan & K. Menou 2011).

$$w_a = \frac{V_e^2 - 2V_d V_e + 2V_d^2}{-V_e + 2V_d}, \qquad (V_e \leqslant 0) \tag{A7a}$$

$$w_s = \frac{2V_d^2}{-V_e + 2V_d}. \tag{A7b}$$

The above equations yield the dimensionless advection factors when there is net evaporation ($V_e > 0$) and deposition ($V_e \leqslant 0$). These parameters are then used to calculate the frictional drag, $\tau$, and sensible heat, $Q_{\text{sens}}$:

$$\tau = -\rho \, V \, w_a, \tag{A8}$$

$$Q_{\text{sens}} = \rho\left(w_s C_p T_s + w_a\left(\frac{V^2}{2} + C_p T\right)\right). \tag{A9}$$

In Equation (A9), the first term is the heat contribution from the surface and the second term is the atmospheric advection. The latent heat term $Q_{\text{lat}}$ is simply:

$$Q_{\text{lat}} = L \, m \, C, \tag{A10}$$

where $L$ is the latent heat of vaporization and $C$ is the condensation rate. This term is defined such that latent heat is added to the atmosphere when condensing but is taken away from the surface when evaporating; this term will subsequently show up in the surface energy balance equation. The last remaining flux term is the pressure gradient force, where we must integrate the vertical pressure with respect to height from a specified altitude to infinity:

$$\int_{z=z_0}^{\infty} P dz = \int_0^{P_0} \frac{\Phi}{g} dP = \frac{C_p}{g} \int_0^{P_0} (T_0 - T) dP. \tag{A11}$$

Equation (A11) switches from the local $z$-altitude coordinates to $P$ levels by substituting the geopotential variable $\Phi$. By expressing $\Phi$ as the work done by moving the air parcel vertically under the gravitational acceleration $g$, we can relate it to enthalpy, which yields a function dependent on the pressure–temperature profile. For an adiabatic profile, completing the vertical integration yields

$$\frac{C_p}{g} \int_0^{P_0} (T_0 - T) \, dP = \frac{C_p \, P_0 \, T_0 \, \beta}{g}, \tag{A12}$$

where $\beta = R/(R + C_p)$. Equation (A12) essentially provides an analytical solution to the pressure integral of a specific profile, but one can generate another temperature profile and solve Equation (A11) numerically.

With the exception of $Q_{\text{RC}}$, Equations (A4), (A8), (A9), and (A12) show how all of the fluxes are calculated in Equations (A1)–(A3). The next step is to convert the local coordinates to global coordinates in terms of $\theta$, the angular distance from the substellar point. Steps detailing the full conversion can be found in A. P. Ingersoll et al. (1985), but the net result is shown:

$$\frac{1}{r \sin(\theta)} \frac{d}{d\theta}\left(\frac{V_* \, P_* \, \sin(\theta)}{g}\right) = m \, E, \tag{A13}$$

$$\frac{1}{r \sin(\theta)} \frac{d}{d\theta}\left(\frac{(V_*^2 + \beta \, C_p \, T_*)P_* \, \sin(\theta)}{g}\right)$$
$$= \frac{\beta \, C_p \, T_* \, P_*}{g \, r \, \tan(\theta)} + \tau, \tag{A14}$$

$$\frac{1}{r\,\sin(\theta)}\frac{d}{d\theta}\left(\frac{(V_*^2/2 + C_p\,T_*)V_*\,P_*\,\sin(\theta)}{g}\right)$$
$$= Q_{\text{sens}} + Q_{\text{lat}} + Q_{\text{RT}}. \quad (A15)$$

The asterisk subscript on the state variables denotes its value at the turbulent boundary layer. Because surface drag induces a logarithmic vertical wind profile, the wind speed at half the scale height is closest to the weighted mean flow of the air column. Therefore, the turbulent boundary layer is taken to be at half the scale height to harmonize our treatment of mass, momentum, and energy advection.

The equations are solved as a set of ordinary differential equations with respect to $\theta$. We use the "shooting" method, where we guess the boundary conditions at the substellar point. Because we are dealing with nonlinear mass, momentum, and energy fluxes, a steady-state solution is particularly sensitive to the initial conditions. The only definitive condition is $V(0) = 0$; for $T(0)$ and $P(0)$, we iteratively have to solve for the conditions that are most numerically stable, as described in T. G. Nguyen et al. (2022). Once a stable solution is established, the flow inevitably reaches supersonic speed and we linearly interpolate the transition from subsonic to supersonic flow.

As the atmosphere gradually condenses out, there comes a point when the atmosphere transitions into an exosphere, where molecules behave less like a continuous fluid and more like independent particles in ballistic motion. As the flow regime changes, our calculations fail. The result is a spatially bounded atmosphere spanning more or less a hemisphere with steady-state evaporation and condensation.

### A.2. Radiative Transfer Formulation

To calculate the radiative contributions to the atmosphere and surface, we first define the vertical optical depth, $\tau_d$, as a function of pressure:

$$\tau_d = x_a P_*/mg, \quad (A16)$$

where $x_a$ is the wavelength-dependent absorption cross section (see Figure 1). The absorptivity/emissivity $\epsilon$ of the overlying atmospheric column is related to the optical depth:

$$\epsilon = 1 - e^{-\tau_d}. \quad (A17)$$

The emissivity $\epsilon$ helps to calculate the individual radiative contribution from stellar radiation absorption ($F_{\text{stel}}$), surface blackbody radiation ($F_{\text{surf}}$), and radiative cooling ($F_{\text{RC}}$). The total heat contribution from radiative transfer, $Q_{\text{RT}}$, becomes:

$$\begin{aligned}Q_{\text{RT}} &= F_{\text{stel}} + F_{\text{surf}} - 2F_{\text{RC}} - F_c \\ &= \int_\lambda \epsilon(\lambda) F_*(\theta, \lambda) d\lambda + \int_\lambda \epsilon(\lambda) B(T_s, \lambda) d\lambda \\ &\quad - 2\int_\lambda \epsilon(\lambda) B(T, \lambda) d\lambda - \frac{1}{2}\epsilon_c\,F_{*,\text{vis}},\end{aligned} \quad (A18)$$

where $F_*$ is the stellar flux and $B$ is the Planck blackbody function, which depends on both wavelength and temperature. Since lava planets are so close to their star, parallel ray approximations no longer hold and $F_*$ is not easily solvable; we use the formulation of W. Kang et al. (2023) to account for finite-light-source effects. Note that the radiative cooling term is doubled, to signify that the atmosphere radiates both down into the ground and outward into space.

Although Equation (A18) is identical to what we had in T. G. Nguyen et al. (2022), barring cloud reflection, we have significantly improved the numerical integration of the stellar flux and blackbody radiation. Previously, we averaged the emissivity across the entire UV and IR wavelength bands, while neglecting wavelengths in the optical completely. The problem is that the absorption cross section of SiO varies enormously with changing temperature and pressure. We therefore calculate absorption and emission line by line with varying pressure and temperature ahead of the simulation. Stellar absorption depends mostly on pressure, and a function of $F_*(P)$ is extracted for easy implementation. Radiative cooling is much more tricky, as it depends on both pressure and temperature, therefore a lookup table of flux is generated (see Figure 7 to quickly determine the radiative transfer terms).

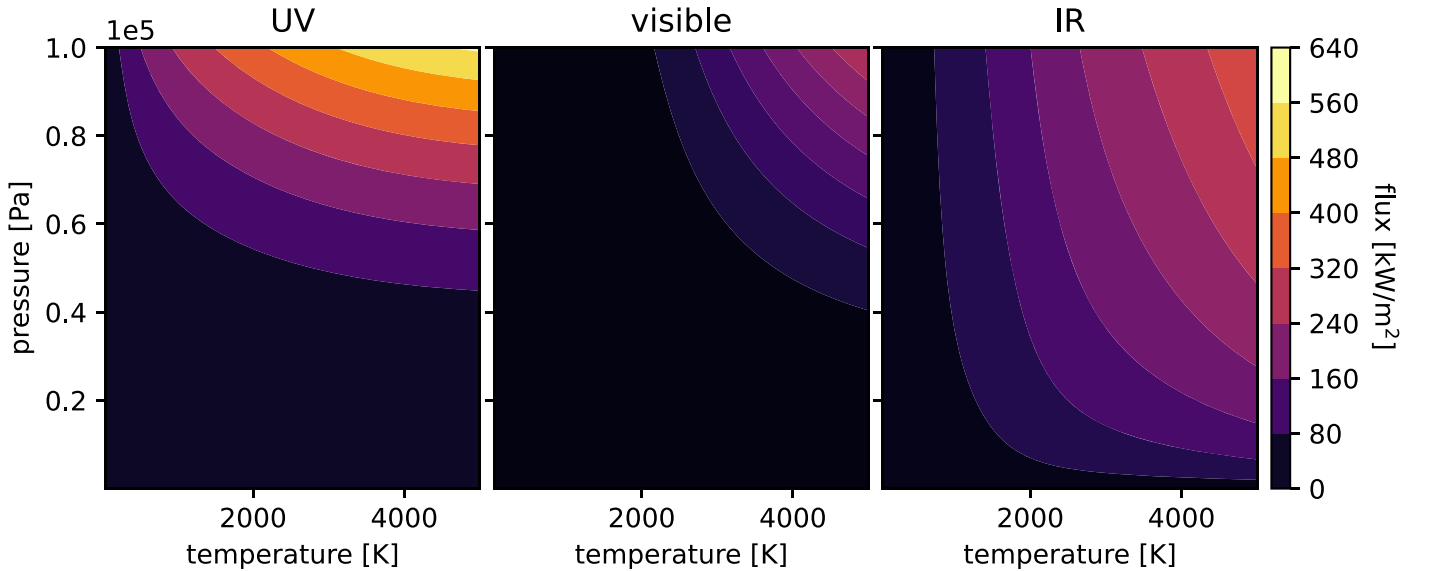

**Figure 7.** Radiative cooling as a function of pressure and temperature: we calculate the radiative cooling and separate the fluxes into three channels: UV (left), visible (middle), and IR (right).

*A.3. Radiative Transfer Validation*

We validate our RT methods with M. Zilinskas et al. (2022), which featured atmospheric modeling for some of the same planet as this work. M. Zilinskas et al. (2022) had more chemical complexities in a silicate-dominant atmosphere while completely neglecting winds, but it is nonetheless useful for comparing equilibrium temperature. Our temperatures at the same pressure levels are within 10% of M. Zilinskas et al. (2022), suggesting our net radiative flux balances are fairly similar. This can be seen in Figure 8.

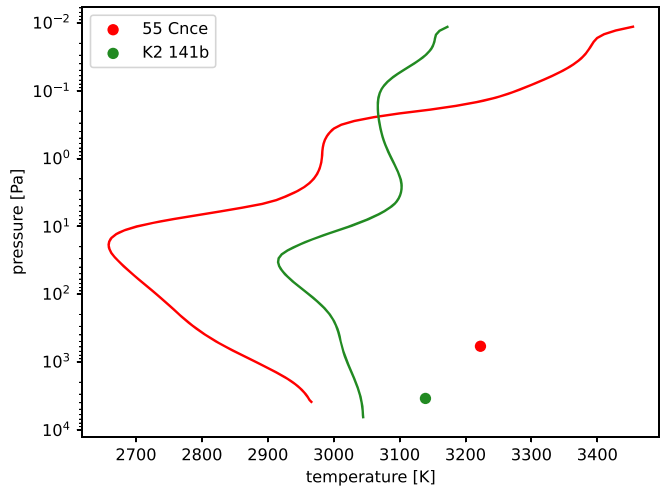

**Figure 8.** Near-surface atmospheric temperature and pressure of our work compared the $T$–$P$ profiles of M. Zilinskas et al. (2022). The $T$–$P$ profiles from M. Zilinskas et al. (2022) are calculated from a representative surface equilibrium temperature of the planet. We plotted our result at the boundary layer matching the same representative surface equilibrium temperature. Our work is ultimately agnostic to the actual shape of the $T$–$P$ profile, as long as the pressure integral up to the surface boundary layer in Equation (A11) matches that of an adiabat. Although the overall dynamics are different, the net radiative flux balances within the bulk of the silicate atmosphere are fairly similar in both studies, thus producing similar atmospheric equilibrium temperatures.

**ORCID iDs**

T. Giang Nguyen ⓘ https://orcid.org/0000-0001-6171-200X
Nicolas B. Cowan ⓘ https://orcid.org/0000-0001-6129-5699
Lisa Dang ⓘ https://orcid.org/0000-0003-4987-6591